\documentclass[%
 aip,
 cha,
 amsmath,amssymb,
 reprint,%
 floatfix,%
]{revtex4-1}

\usepackage{graphicx}
\usepackage{dcolumn}
\usepackage{bm}
\usepackage{multirow}

\usepackage[utf8]{inputenc}
\usepackage[T1]{fontenc}
\usepackage{mathptmx}
\usepackage{etoolbox}

\AtBeginDocument{\hfuzz=30pt}

\makeatletter
\let\frontmatter@correspondingemails\@empty
\def\@email#1#2{%
 \endgroup
 \g@addto@macro\frontmatter@correspondingemails{\produce@RRAP{#1\href{mailto:#2}{#2}}\par}%
}%
\AtBeginDocument{%
 \patchcmd{\titleblock@produce}
  {\frontmatter@RRAPformat}
  {\frontmatter@RRAPformat{\frontmatter@correspondingemails}\frontmatter@RRAPformat}
  {}{}
}%
\makeatother
\begin{document}

\preprint{AIP/123-QED}

\title{Plasma Instabilities in Arbitrary Distributions: Comparison between ALPS and BO}
\author{Xudong Guo}
\affiliation{%
Purple Mountain Observatory, Chinese Academy of Sciences, Nanjing 210023, China
}%
\affiliation{%
School of Astronomy and Space Science, University of Science and Technology of China, Hefei 230026, China
}%

\author{Huasheng Xie$^{*}$}
\affiliation{%
Beijing VeloAlpha Technology Co., Ltd., Beijing, 100080, China
}%
\email[Corresponding author: ]{huashengxie@gmail.com}

\author{Kristopher G. Klein}
\affiliation{%
Lunar and Planetary Laboratory, University of Arizona, Tucson, AZ 85721, USA
}%

\author{D. Verscharen}
\affiliation{%
Mullard Space Science Laboratory, University College London, Dorking RH5 6NT, UK
}%

\author{Chen Shi}
\affiliation{%
Changsha University of Science and Technology, Changsha, China
}%

\author{Jinsong Zhao$^{*}$}
\affiliation{%
Purple Mountain Observatory, Chinese Academy of Sciences, Nanjing 210023, China
}%
\email[Corresponding author: ]{zhaojs82@gmail.com}

\date{\today}

\begin{abstract}

Determining accurate wave dispersion relations is a central problem in plasma physics. Recent advances have enabled the numerical computation of linear dispersion relation in plasmas with arbitrary particle velocity distribution functions (VDFs), using two distinct solvers, BO and ALPS. Their reliability and mutual consistency, however, have not been systematically tested for a broad range of VDFs. Here we compare the dispersion relations obtained from BO and ALPS for several representative distributions. We find that the two solvers give consistent unstable modes for kappa distributions with large values of $\kappa$, as well as for ring-beam, shell, and proton core-beam distributions. BO, however, becomes unreliable for kappa distributions with $\kappa < 4$. For an observationally derived VDF, the two solvers give similar real frequencies for the unstable waves but substantially different growth rates. This difference is mainly caused by the imperfect fitting of the input distribution required by BO. Despite this limitation, BO has a clear computational advantage because it can obtain all roots in a single run. Considering the complementary strengths of the two solvers, their combined use can provide a more reliable and effective framework for investigating instabilities in non-Maxwellian plasma environments.

\end{abstract}

\maketitle

\section{Introduction}

Understanding wave modes in astrophysical, space, and laboratory plasmas relies on determining their dispersion relations \cite{stix1992waves}, which describe the relationship between wave frequency and wave vector. Although analytical dispersion relations can be derived from fluid and kinetic theories \cite{stix1992waves,zhao2015,yoon2026PhPl}, they generally depend on simplifying assumptions, such as restricted frequency ranges or idealized propagation geometries. These limitations hinder their applicability to realistic plasma environments, motivating the development of numerical approaches that can describe plasma waves and instabilities under more general conditions \cite{ronnmark1982whamp,Astfalk2015,Verscharen_2018,verscharen2018alps,xie2014,xie2016pdrk,xie2019bo,bai2025bo}.

A key challenge in such numerical approaches is the efficient and reliable solution of the dispersion relation. Existing solvers generally adopt two distinct strategies. The first treats the problem as a root-finding task in the complex frequency plane, as implemented in solvers such as WHAMP \cite{ronnmark1982whamp}, DSHARK \cite{Astfalk2015}, NHDS \cite{Verscharen_2018}, ALPS \cite{verscharen2018alps,klein2025dielectric}, and PLUME \cite{Klein_2025}, which typically yield one solution for a given wave vector in each calculation. The second reformulates the problem as a matrix eigenvalue problem, as in BO \cite{xie2014,xie2016pdrk,xie2019bo,bai2025bo}, enabling the simultaneous computation of all wave modes at a given wave vector. These approaches have been extensively applied to plasmas with pre-defined particle velocity distribution functions (VDFs), such as drifting Maxwellian \cite{sun2019kinetic,sun2020electron,sun2021electron} and kappa distributions \cite{astfalk2017leopard,bai2025bo,Schroder2025ApJ}. 

Recently, both types of solvers have been extended to arbitrary gyrotropic VDFs, as implemented in ALPS \cite{verscharen2018alps,klein2025dielectric} and BO \cite{xie2025efficient}. This extension is important because observed plasma distributions often differ substantially from pre-defined and idealized forms such as Maxwellian or kappa distributions. ALPS and BO have already been applied to several problems involving non-standard VDFs \cite{klein2025dielectric,xie2025efficient,klein2026GeoRL,10.1063/5.0260878,Tischmann_2026,Zhang_2025}. However, their reliability and mutual consistency have not yet been examined systematically for a broad range of VDFs.

In this study, we compare the dispersion relations obtained from BO and ALPS for six representative VDFs with analytical expressions and one observationally derived VDF. The analytical cases allow us to test the two solvers under controlled conditions. We find that BO and ALPS give consistent instability results in most cases, but BO becomes unreliable for kappa distributions with small kappa indices. For the observationally derived VDF, the two solvers produce similar real frequencies for unstable waves but noticeably different growth rates. The discrepancy mainly arises from the distribution fitting required by BO. These results clarify the regimes in which the two solvers can be used reliably and show that BO and ALPS are best regarded as complementary tools for studying instabilities in plasmas with non-Maxwellian VDFs.

\section{Two Solver Strategies for Finding Dispersion Relations}

\subsection{BO: Matrix Eigenvalue Solver}

BO is implemented in MATLAB, and a Julia version is also available. The core idea of the BO solver is to transform the kinetic dispersion relation of a magnetized plasma into a standard matrix eigenvalue problem through a rational approximation of the plasma dispersion function. This approach avoids the initial-guess dependence required by traditional point-by-point root searching in the complex-frequency domain and enables all major wave-mode solutions to be obtained in a single run.

Solving the dispersion relation requires the evaluation of the conductivity tensor $\sigma$. For an arbitrary non-relativistic gyrotropic plasma distribution, the conductivity tensor is written as:
\begin{equation}
\sigma=-i\sum_{s}\frac{q_{s}^2 n_{\mathrm{s0}}}{m_{s}}\sum_{n=-\infty}^{\infty}\int_{-\infty}^{\infty}\int_0^{\infty}\frac{2\pi v_\perp\,dv_\perp\,dv_\parallel}{\omega-n\omega_{\mathrm{cs}}-k_\parallel v_\parallel}\,\Pi_{s},
\end{equation}
with 
\begin{equation}
\Pi_{s}=\left[
\begin{array}{ccc}
A_{s}\dfrac{n^2 v_\perp}{\mu_{s}^2}J_{n}^2 & iA_{s}\dfrac{n v_\perp}{\mu_{s}}J_{n}J_{n}' & B_{s}\dfrac{n v_\perp}{\mu_{s}}J_{n}^2 \\
-iA_{s}\dfrac{n v_\perp}{\mu_{s}}J_{n}J_{n}' & A_{s} v_\perp J_{n}'^2 & -iB_{s} v_\perp J_{n}J_{n}' \\
A_{s}\dfrac{n v_\parallel}{\mu_{s}}J_{n}^2 & iA_{s} v_\parallel J_{n}J_{n}' & B_{s} v_\parallel J_{n}^2
\end{array}
\right],
\end{equation}
where
\begin{equation*}
\begin{aligned}
\mu_{s}&=\frac{k_\perp v_\perp}{\omega_{\mathrm{cs}}},\\
A_{s}&=\left(1-\frac{k_\parallel v_\parallel}{\omega}\right)\frac{\partial f_{\mathrm{s0}}}{\partial v_\perp}+\frac{k_\parallel v_\perp}{\omega}\frac{\partial f_{\mathrm{s0}}}{\partial v_\parallel},\\
B_{s}&=\frac{n\omega_{\mathrm{cs}}v_\parallel}{\omega v_\perp}\frac{\partial f_{\mathrm{s0}}}{\partial v_\perp}+\left(1-\frac{n\omega_{\mathrm{cs}}}{\omega}\right)\frac{\partial f_{\mathrm{s0}}}{\partial v_\parallel}.
\end{aligned}
\end{equation*}
Here, $J_{n}\equiv J_{n}(\mu_{s})$ is the Bessel function of the first kind of order $n$, and $J_{n}'\equiv dJ_{n}(\mu_{s})/d\mu_{s}$. To evaluate the distribution-function gradients in the parallel and perpendicular directions, BO represents arbitrary VDFs using orthogonal basis-function expansions. In this work, we use a Hermite-basis (HH) expansion for this purpose. 

For the HH expansion, we use
\begin{equation}
f_{\mathrm{s0}}(v_\parallel,v_\perp)=c_{\mathrm{s0}}\sum_{l=0}^{\infty}\sum_{m=0}^{\infty}a_{s,\mathrm{lm}}\,g_{\mathrm{sz},l}(v_\parallel)\,g_{\mathrm{sx},m}(v_\perp),
\end{equation}
with
\begin{equation*}
\begin{aligned}
&g_{\mathrm{sz},l}(v_\parallel)=\left(\frac{v_\parallel-d_{\mathrm{sz}}}{L_{\mathrm{sz}}}\right)^l e^{-\left(\frac{v_\parallel-d_{\mathrm{sz}}}{L_{\mathrm{sz}}}\right)^2},\\
&g_{\mathrm{sx},m}(v_\perp)=\left(\frac{v_\perp-d_{\mathrm{sx}}}{L_{\mathrm{sx}}}\right)^m e^{-\left(\frac{v_\perp-d_{\mathrm{sx}}}{L_{\mathrm{sx}}}\right)^2}.
\end{aligned}
\end{equation*}
Here, $L_{\mathrm{sz}}$ and $L_{\mathrm{sx}}$ are velocity widths, and $d_{\mathrm{sz}}$ and $d_{\mathrm{sx}}$ are drift velocities. The normalization coefficient is $c_{\mathrm{s0}}=1/(\pi^{3/2}L_{\mathrm{sz}}L_{\mathrm{sx}}^2R_{s})$, with $R_{s}=\exp[-d_{\mathrm{sx}}^2/(L_{\mathrm{sx}}^2)]+\sqrt{\pi}d_{\mathrm{sx}}/L_{\mathrm{sx}}\,\mathrm{erfc}\!\left(-d_{\mathrm{sx}}/L_{\mathrm{sx}}\right)$, and $\mathrm{erfc}(-x)=1-\mathrm{erf}(-x)=1+\mathrm{erf}(x)$ is the complementary error function. The coefficients $a_{s,\mathrm{lm}}$ are obtained from orthogonal Hermite-basis projections. BO define
\begin{equation*}
\begin{aligned}
&f_{\mathrm{s0},\mathrm{lm}}(v_{\parallel},v_{\perp})\equiv a_{s,\mathrm{lm}}g_{\mathrm{sz},l}(v_{\parallel})g_{\mathrm{sx},m}(v_{\perp}),\\
&f_{\mathrm{s0z},l}(v_{\parallel})\equiv g_{\mathrm{sz},l}(v_{\parallel}),\\
&f_{\mathrm{s0x},m}(v_{\perp})\equiv g_{\mathrm{sx},m}(v_{\perp}).
\end{aligned}
\end{equation*}
 
 Accordingly, the derivatives are given by $\partial f_{\mathrm{s0z},l}(v_{\parallel})/\partial v_{\parallel}=-(1/L_{\mathrm{sz}})\left[2f_{\mathrm{s0z},l+1}-l f_{\mathrm{s0z},l-1}\right]$ and $\partial f_{\mathrm{s0x},m}(v_{\perp})/\partial v_{\perp}=-(1/L_{\mathrm{sx}})\left[2f_{\mathrm{s0x},m+1}-m f_{\mathrm{s0x},m-1}\right]$. Thus, the distribution gradients are transformed into sums of adjacent terms.

 For the parallel-direction integral, we define
\begin{equation*}
\begin{aligned}
&Z_{l,p}(\zeta_{\mathrm{sn}})\equiv -\frac{k_\parallel}{L_{\mathrm{sz}}^{p}\sqrt{\pi}}\int_{-\infty}^{\infty}\frac{v_\parallel^{p}\left(\frac{v_\parallel-d_{\mathrm{sz}}}{L_{\mathrm{sz}}}\right)^{l}e^{-\left(\frac{v_\parallel-d_{\mathrm{sz}}}{L_{\mathrm{sz}}}\right)^{2}}}{\omega-k_\parallel v_\parallel-n\omega_{\mathrm{cs}}}\,dv_\parallel\\
&=\frac{1}{\sqrt{\pi}}\int_{-\infty}^{\infty}\frac{g_{l,p}(z)}{z-\zeta_{\mathrm{sn}}}\,dz,\\
&g_{l,p}(z)\equiv(z+d)^{p}z^{l}e^{-z^{2}}.
\end{aligned}
\end{equation*}
where $z\equiv(v_\parallel-d_{\mathrm{sz}})/L_{\mathrm{sz}}$ and $d\equiv d_{\mathrm{sz}}/L_{\mathrm{sz}}$. We have $Z_{0,0}(\zeta)=Z(\zeta)$, and redefine $Z_{l,0}\equiv Z_{l}$ and $I_{l}\equiv \pi^{-1/2}\int_{-\infty}^{\infty}x^{l}\exp[-x^2]\,dx$, with $\zeta_{\mathrm{sn}}=(\omega-k_\parallel d_{\mathrm{sz}}-n\omega_{\mathrm{cs}})/(k_\parallel L_{\mathrm{sz}})$, $d=d_{\mathrm{sz}}/L_{\mathrm{sz}}$, and $p=0,1,2$. These plasma dispersion functions can be computed efficiently to high accuracy and analytic continuation extends the functions to $\operatorname{Im}(\zeta_{\mathrm{sn}})\le 0$.

For the perpendicular integral, we define
\begin{equation*}
\begin{array}{l}
\Gamma_{\{a,b,c\}}^{n,m,p}(a_{s},d_{s}) \\
= \dfrac{1}{L_{\mathrm{sx}}^{p+1}}\int_0^{\infty} v_\perp^{p}
\left\{J_{n}^2\!\left(\frac{k_\perp v_\perp}{\omega_{\mathrm{cs}}}\right),\,
J_{n}\!\left(\frac{k_\perp v_\perp}{\omega_{\mathrm{cs}}}\right)J_{n}'\!\left(\frac{k_\perp v_\perp}{\omega_{\mathrm{cs}}}\right),\,
J_{n}'^2\!\left(\frac{k_\perp v_\perp}{\omega_{\mathrm{cs}}}\right)\right\} \\
\times \left(\frac{v_\perp-d_{\mathrm{sx}}}{L_{\mathrm{sx}}}\right)^m
e^{-\left(\frac{v_\perp-d_{\mathrm{sx}}}{L_{\mathrm{sx}}}\right)^2}dv_\perp \\
= \int_0^{\infty} x^{p}\left\{J_{n}^2(a_{s}x),\,J_{n}(a_{s}x)J_{n}'(a_{s}x),\right.
\left.\,J_{n}'^2(a_{s}x)\right\}(x-d_{s})^m e^{-(x-d_{s})^2}\,dx,
\end{array}
\end{equation*}
with $a_{s}\equiv k_\perp L_{\mathrm{sx}}/\omega_{\mathrm{cs}}$ and $d_{s}\equiv d_{\mathrm{sx}}/L_{\mathrm{sx}}$. The most critical step in BO is the rational expansion of the $Z$ function:
\begin{equation}
Z_{l}(\zeta) \simeq \sum_{j=1}^{J}\frac{b_{j} c_{j}^{l}}{\zeta-c_{j}}.
\end{equation}
Finally, $\sigma$ can be written in compact matrix form as
\begin{equation}
-\frac{\sigma}{i\epsilon_0}=\begin{pmatrix}
M_{11} & M_{12} & M_{13} \\
M_{21} & M_{22} & M_{23} \\
M_{31} & M_{32} & M_{33}
\end{pmatrix},
\end{equation}
with
\begin{equation}
M_{\mathrm{ij}}=\frac{b_{\mathrm{ij}}}{\omega}+\sum_{\mathrm{snj}}\frac{b_{\mathrm{snj,ij}}}{\omega-c_{\mathrm{snj}}}.
\end{equation}
By combining $\mathbf{J}=\sigma\cdot\mathbf{E}$ with Maxwell's equations, solving the dispersion relation is transformed into a matrix eigenvalue problem. Instead of returning to the traditional determinant root-finding form, BO solves the dispersion relation through an equivalent coupled Ohm's-law--Maxwell system; the detailed derivation is given in Ref.~\onlinecite{xie2025efficient}. 

\subsection{ALPS: Iterative Root-finding Solver}
ALPS (Arbitrary Linear Plasma Solver), written in Fortran, adopts the strategy of numerically constructing the dispersion tensor directly from an arbitrary prescribed background distribution and solving the kinetic dispersion relation in the complex-frequency plane. ALPS is formulated in momentum space, which allows a straightforward extension to relativistic plasmas. In the present work, however, we focus exclusively on non-relativistic solutions. For a given background magnetic field $\mathbf{B}_0=(0,0,B_0)$ and wave vector $\mathbf{k}=(k_\perp,0,k_\parallel)$, the linear fluctuations satisfy $\det\!\left[\mathbf{D}(\omega,\mathbf{k})\right]=0$. Unlike BO, ALPS directly computes the linear-response tensor numerically on a discrete velocity grid. In the non-relativistic setting considered here, its basic input is a tabulated gyrotropic distribution function $f_{\mathrm{s0}}(v_{\parallel},v_{\perp})$, and the derivatives $\partial f_{\mathrm{s0}}/\partial v_{\parallel}$ and $\partial f_{\mathrm{s0}}/\partial v_{\perp}$ are evaluated on the same grid. In addition, ALPS allows any selected species to be specified as a bi-Maxwellian distribution instead of a tabulated arbitrary distribution, in which case the corresponding susceptibility is evaluated using the analytic expressions implemented in NHDS \cite{Verscharen_2018}; thus Maxwellian and arbitrary-distribution species can be combined within a single calculation.

The species response can be written in a unified form as a two-dimensional integral in velocity space:
\begin{equation}
\chi_{s}(\omega,\mathbf{k}) \propto \sum_{n=-\infty}^{+\infty}\int_0^{\infty}2\pi v_\perp\,dv_\perp\int_{-\infty}^{+\infty}dv_\parallel\,\frac{\mathbf{T}_{\mathrm{sn}}(v_\parallel,v_\perp)}{\omega-k_\parallel v_\parallel-n\Omega_{s}}.
\end{equation}
Here, $\Omega_{s}$ is the cyclotron frequency of species $s$, and $\mathbf{T}_{\mathrm{sn}}$ is the tensor kernel built from the distribution function, its velocity derivatives, and Bessel terms. 

In ALPS, the susceptibility is obtained by numerically evaluating this two-dimensional integral for each species and harmonic $n$, and then constructing the full dispersion tensor $\mathbf{D}(\omega,\mathbf{k})$. ALPS treats the two integrations differently: the perpendicular part is kept as a standard real-axis integral, and for each fixed $v_\perp$ the parallel integral is evaluated first and then integrated over $v_\perp$. Thus, the two-dimensional integral can be written as
\begin{equation}
\chi_{s}(\omega,\mathbf{k}) \propto \sum_{n} \int_0^{\infty} 2\pi v_\perp\,dv_\perp\,I_{\mathrm{sn}}(v_\perp),
\end{equation}
where
\begin{equation}
I_{\mathrm{sn}}(v_\perp)=\int_{-\infty}^{+\infty}dv_\parallel\,\frac{\mathbf{T}_{\mathrm{sn}}(v_\parallel,v_\perp)}{\omega-k_\parallel v_\parallel-n\Omega_{s}}.
\end{equation}

Parallel integration is a central step in ALPS. For $\gamma\le 0$, the Landau-contour integral must include pole contributions in the complex $v_\parallel$ plane due to the analytic continuation. The integral is therefore written as a real-axis principal part plus pole-residue terms. ALPS addresses this with a hybrid analytic-continuation method: for each fixed $v_\perp$, it builds a representation of the slice $f_{\mathrm{s0}}(v_\parallel,v_\perp=\mathrm{const})$ that can be evaluated in the complex plane, and then evaluates $f_{\mathrm{s0}}$ and its parallel derivative at the complex resonance point $v_{\mathrm{res}}$. The first version of ALPS used fitting of each slice with a prescribed analytic distribution; in the code, this option corresponds to $\mathrm{AC}=1$. The improved ALPS offers the option to use a Chebyshev/GLLS polynomial representation for each slice, corresponding to $\mathrm{AC}=2$, which converts discrete real-axis data into a function $\tilde{f}_{\mathrm{s0}}(z, v_\perp)$ that can be evaluated in the complex $v_\parallel$ plane \cite{klein2025dielectric}. Values at $z=v_{\mathrm{pole}}$ are then used to compute pole contributions for damped modes.

In the root-finding stage, ALPS employs an iterative Newton--secant algorithm that updates $\omega_{r}$ and $\gamma$ simultaneously at fixed wave vector $\mathbf{k}$. The algorithm requires an initial guess; in practice, this is often obtained from local minima in the dispersion map of $\log | \det \mathbf{D}(\omega_r, \gamma)|$. The updated values of $\omega_{r}$ and $\gamma$ are then fed back into the numerical evaluation of the susceptibility tensor, and the iteration is repeated until convergence.

\section{Instability Comparison Across Six Representative VDFs}\label{sec3}



This section compares the instabilities obtained with ALPS and BO for six representative VDFs defined by analytical expressions. The baseline plasma parameters for all six cases, including the number density $n_s$, parallel and perpendicular drift velocities $v_{\mathrm{d}\parallel s}$ and $v_{\mathrm{d}\perp s}$, and parallel and perpendicular temperatures $T_{s\parallel}$ and $T_{s\perp}$ for species $s$, are summarized in Table~\ref{tab:particle_all_cases} in Appendix~\ref{app_parameters}. Since the instability results can be sensitive to the VDF fitting, we present the fitting results for the kappa distributions in Case 1 below and provide the corresponding results for the other distributions in Appendix~\ref{fit_com}.

\subsection{Case 1: Oblique Proton Firehose Instability Driven by a Proton Kappa Distribution}

\begin{figure}[t!]
\centering
\includegraphics[width=0.95\columnwidth]{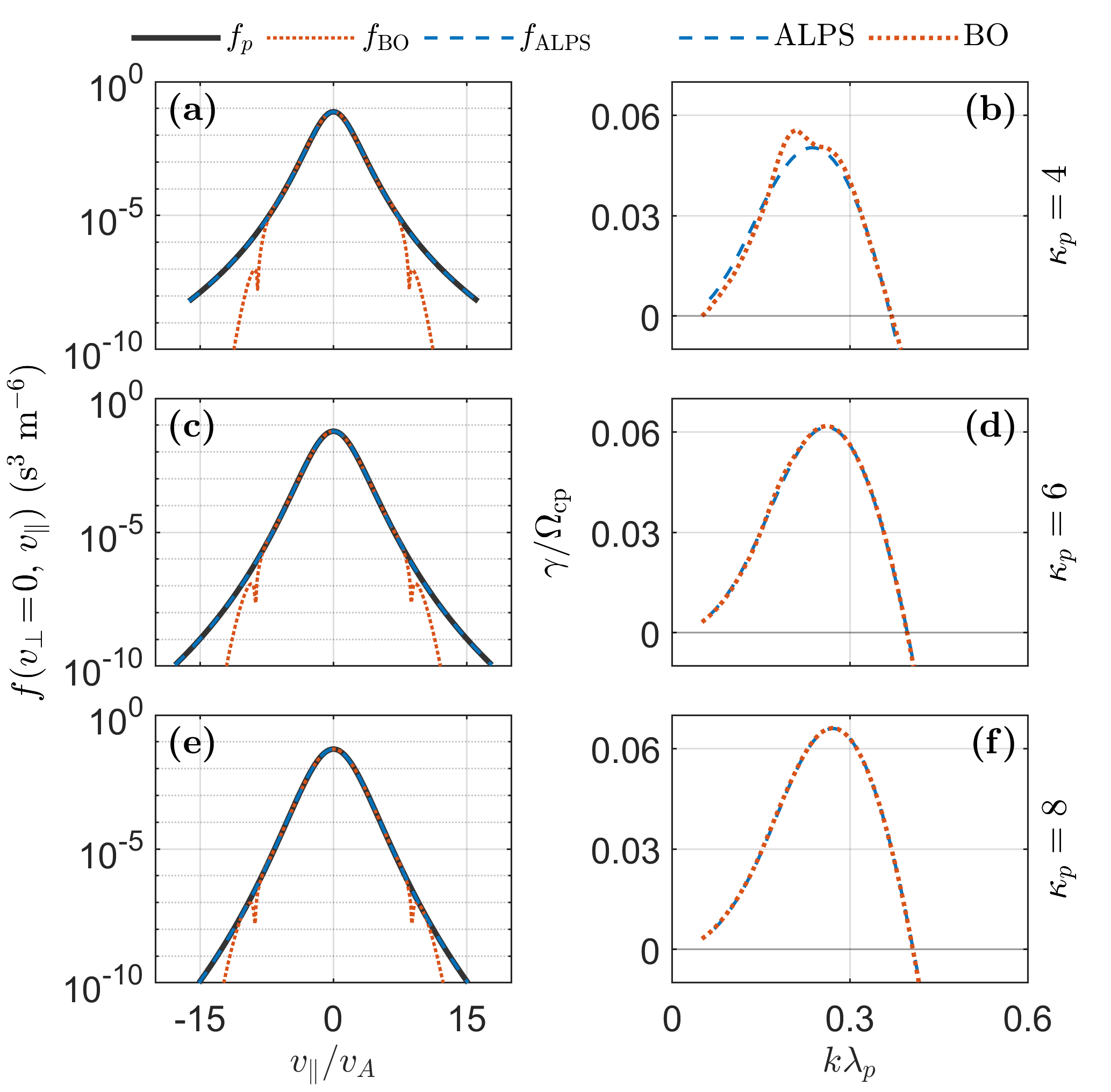}
\caption{Case 1: Comparison of fits to the proton VDF $f_p$ and the resulting instability growth rates $\gamma$ obtained using the BO (red dashed curves) and ALPS (blue dotted curves) solvers. Results are shown for three proton kappa indices: (a, b) $\kappa_p=4$; (c, d) $\kappa_p=6$; and (e, f) $\kappa_p=8$. The proton and electron populations are described by a kappa distribution and a Maxwellian distribution, respectively. Their analytical expressions are given in Eqs.~(\ref{fp_case1}) and (\ref{fe_case1}).}
\label{fig:kappa_case}
\end{figure}

In Case 1, we examine the oblique proton firehose instability driven by a proton kappa distribution with anisotropic temperatures, while the electrons are assumed to follow a Maxwellian distribution with isotropic temperatures. 

The proton kappa distribution and the electron Maxwellian distribution are defined as: 
\begin{eqnarray}
\label{fp_case1}
& f_{p} = \frac{n_p}{\pi^{3/2}c_\mathrm{p\parallel}c_\mathrm{p\perp}^{2}} 
           \frac{\kappa_{p}^{3/2}\Gamma\!\left(\kappa_{p}-\tfrac{1}{2}\right)}{\Gamma(\kappa_{p}+1)}
 \left[1+\frac{(v_{\parallel}-v_\mathrm{d\parallel p})^{2}}{\kappa_{p}c_\mathrm{p\parallel}^{2}}+\frac{v_{\perp}^{2}}{\kappa_{p}c_\mathrm{p\perp}^{2}}\right]^{-\kappa_{p}-1} &
\end{eqnarray}
and
\begin{equation}
\label{fe_case1}
f_{e} = \frac{n_e}{\pi^{3/2}c_\mathrm{e\parallel}c_\mathrm{e\perp}^2}
\exp\!\left[-\frac{ \left(v_{\parallel} - v_\mathrm{d\parallel e} \right)^{2}}{c_\mathrm{e\parallel}^{2}}-\frac{v_{\perp}^{2}}{c_\mathrm{e\perp}^{2}}\right],
\end{equation}
respectively. Here, the proton thermal speeds $c_{\mathrm{p}\parallel}$ and $c_{\mathrm{p}\perp}$ in the kappa distribution are related to the temperatures by:
\begin{equation}
\resizebox{\linewidth}{!}{$
c_{\mathrm{p}\parallel}
=\sqrt{\frac{2k_{B}T_{\mathrm{p}\parallel}}{m_{p}}
\left(1-\frac{3}{2\kappa_{p}}\right)}
\quad \text{and} \quad
c_{\mathrm{p}\perp}
=\sqrt{\frac{2k_{B}T_{\mathrm{p}\perp}}{m_{p}}
\left(1-\frac{3}{2\kappa_{p}}\right)} .
$}
\nonumber
\end{equation}
In contrast, the electron Maxwellian distribution uses $c_\mathrm{e\parallel} = \sqrt{ {2k_{B}T_\mathrm{e\parallel}}/{m_{e}} }$ and $c_\mathrm{e\perp} = \sqrt{{2k_{B}T_\mathrm{e\perp}}/{m_{p}} }$. Additionally, the instability analysis assumes a background magnetic field of $B_0=0.1$~T and a wave propagation angle of $\theta=45^\circ$.



To evaluate the instabilities with the two solvers, we fit the proton VDF $f_p$ using the $\mathrm{AC}=1$ formulation in ALPS and the HH method in BO. Figure~\ref{fig:kappa_case}(a), (c), and (e) show the fitting results from both solvers for three representative kappa indices ($\kappa_p = 4$, $6$, and $8$). Compared with the analytical VDF, the ALPS formulation achieves significantly higher accuracy, whereas the BO fit exhibits pronounced deviations at large parallel velocities.


Figures~\ref{fig:kappa_case}(b), (d), and (f) present the instability growth rates for the three kappa indices. At the lowest kappa index ($\kappa_p = 4$), the growth rate curve from BO deviates considerably from the ALPS result. This discrepancy results from the reduced fitting accuracy of BO in the distribution tails, indicating that BO has difficulty resolving the instability when the kappa index is small. In contrast, for the higher kappa values ($\kappa_p = 6$ and $8$), the two solvers yield highly consistent growth rates.

\subsection{Case 2: Parallel Whistler Instability Driven by an Electron Product-Kappa Distribution}

\begin{figure}[t!]
\centering
\includegraphics[width=0.95\columnwidth]{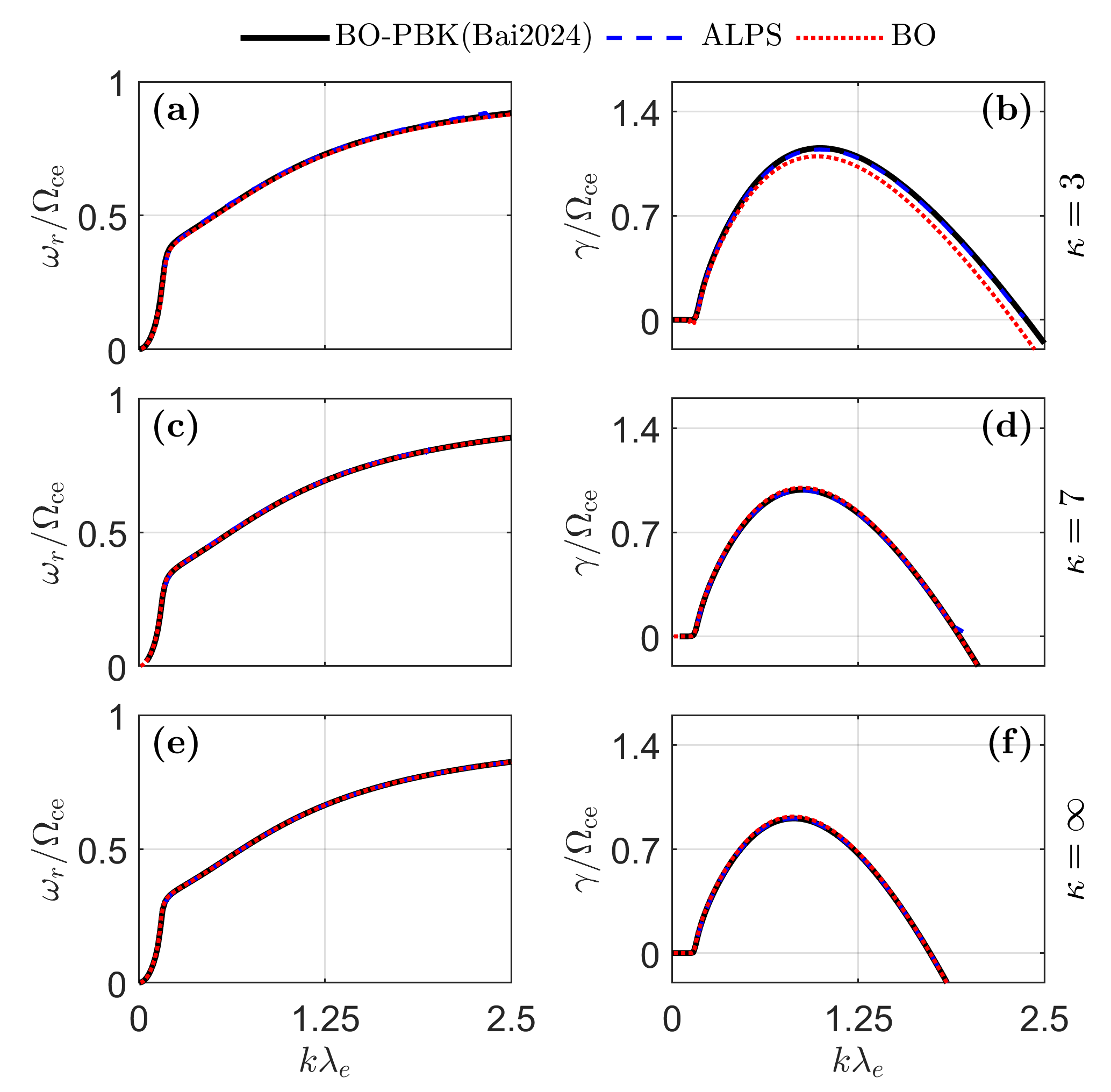}
\caption{Case 2: Comparison of the real frequency $\omega_r$ (left panels) and the growth rate $\gamma$ (right panels) of the parallel whistler instability calculated via BO (red dashed curves) and ALPS (blue dotted curves). Three scenarios are shown: (a, b) $\kappa=3$; (c, d) $\kappa=7$; and (e, f) the Maxwellian limit ($\kappa \to \infty$). Benchmark results from Ref.~\onlinecite{bai2024new} are plotted as black solid curves. The analytical expression for product-kappa distribution is given in Eq.~(\ref{fe_case2}).
}
\label{fig:product_kappa_case}
\end{figure}

In Case 2, we analyze the parallel whistler instability driven by an electron product-kappa distribution with anisotropic temperatures. Instabilities driven by such distributions have attracted significant attention in previous studies \cite{Lazar2015AA,bai2024new,bai2025bo}. Specifically, Ref.~\onlinecite{bai2024new} performed a parameter study on this instability, which serves as an excellent benchmark for the results obtained from BO and ALPS.

The electron product-kappa distribution is given by:
\begin{equation}
\label{fe_case2}
\begin{aligned}
f_{e} &=\frac{1}{\pi^{2}c_\mathrm{e\parallel}c_\mathrm{e\perp}^{2}}
            \frac{\kappa_{\parallel,e}^{1/2}\Gamma\!\left(\kappa_{\parallel,e}+\tfrac{1}{2}\right)}{\Gamma(\kappa_{\parallel,e}+1)} \\
&\quad \times \left[1+\frac{v_{\parallel}^{2}}{\kappa_{\parallel,e}c_\mathrm{e\parallel}^{2}}\right]^{-\kappa_{\parallel,e}-1} 
\left[1+\frac{v_{\perp}^{2}}{\kappa_{\perp,e}c_\mathrm{e\perp}^{2}}\right]^{-\kappa_{\perp,e}-1},
\end{aligned}
\end{equation}
where the electron thermal speeds $c_\mathrm{e\parallel}$ and $c_\mathrm{e\perp}$ are defined by
\begin{equation}
\resizebox{\linewidth}{!}{$
c_\mathrm{e\parallel}
=\sqrt{\frac{2k_{B}T_{z,e}}{m_{e}}
\left(1-\frac{1}{2\kappa_{\parallel,e}}\right)}
\quad \text{and} \quad
c_\mathrm{e\perp}
=\sqrt{\frac{2k_{B}T_{p,e}}{m_{e}}
\left(1-\frac{1}{\kappa_{\perp,e}}\right)} .
$}
\nonumber
\end{equation}

The proton VDF in Case~2 is also assumed to be a product-kappa distribution. Its analytical expression is identical to that in Eq.~(\ref{fe_case2}), but with the subscript ``$p$'' instead of ``$e$''. The background plasma parameters are summarized in Table~\ref{tab:particle_all_cases}, and a magnetic field magnitude of $B_0=1.0\times10^{-8}$~T is used in the instability analysis. Additionally, we apply the constraint $\kappa_{e\parallel}=\kappa_{e\perp}=\kappa_{p\parallel}=\kappa_{p\perp} = \kappa$.

We utilize the $\mathrm{AC}=2$ formulation in ALPS and the HH method in BO to fit the electron VDF. Similar to Case 1, the BO fit exhibits lower accuracy at low kappa indices, whereas ALPS maintains highly consistent fits across all distributions. 


A comparison of the parallel whistler instability calculated by the two solvers is presented in Figure~\ref{fig:product_kappa_case} for three kappa-index scenarios: $\kappa=3$, $7$, and the Maxwellian limit ($\kappa \to \infty$). For the real frequency, both solvers yield results in excellent agreement with Ref.~\onlinecite{bai2024new}. For the growth rate, however, BO gives inconsistent results at the lowest index ($\kappa=3$) and converges with the benchmark only at higher kappa values. In contrast, ALPS accurately captures the growth rate across all three kappa cases. It should be noted, however, that dedicated BO versions have also been developed for kappa and product-kappa distributions \cite{bai2025bo,bai2024new,bai2023toward}. These versions have shown very good performance in a large number of tests and can handle the low-$\kappa$ regime well.

\subsection{Case 3: Oblique Instability Driven by an Electron Ring-Beam Distribution}

\begin{figure}[t!]
\centering
\includegraphics[width=1\columnwidth]{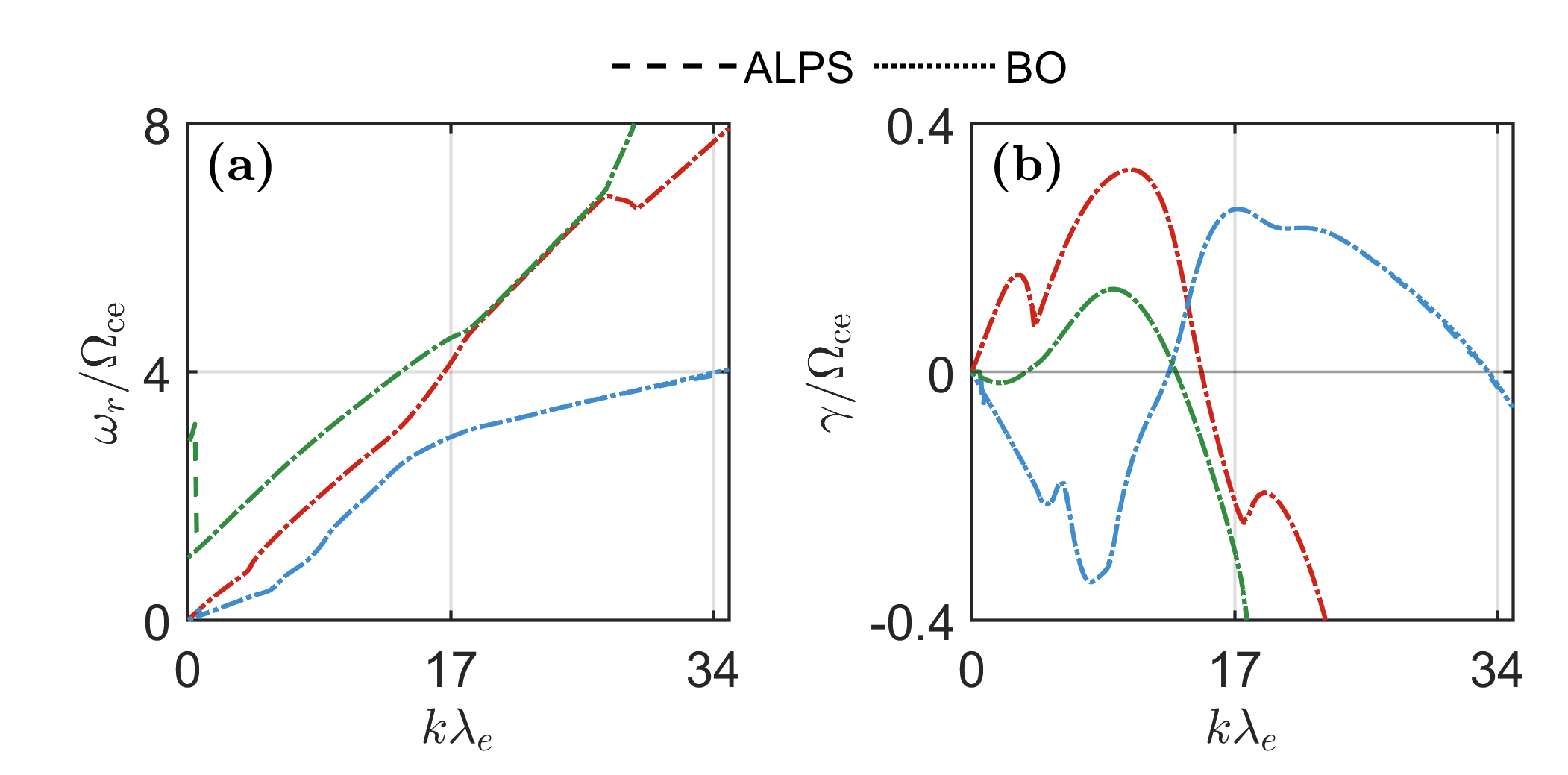}
\caption{Case 3: Comparison of the oblique instability driven by an electron ring-beam distribution. (a) Real frequency $\omega_r$ and (b) growth rate $\gamma$. The analytical expression for the electron ring-beam component is given in Eq.~(\ref{fe_case3}).}
\label{fig:ring_beam_case}
\end{figure}

In Case 3, we consider an oblique instability at $\theta=40^\circ$ driven by an electron ring-beam distribution, given by the following analytical expression:
\begin{equation}
\label{fe_case3}
 f_{e} = \frac{1}{\pi^{3/2}c_\mathrm{e\parallel}c_\mathrm{e\perp}^2A_{e}}
\exp\!\left[-\frac{(v_\parallel-v_{\mathrm{dz},e})^2}{c_\mathrm{e\parallel}^2}
-\frac{(v_\perp-v_{\mathrm{dr},e})^2}{c_\mathrm{e\perp}^2}\right], 
\end{equation}
where $A_{e}= \mathrm{exp}\left( -v_{\mathrm{dr},e}^2/c_\mathrm{e\perp}^2 \right)+\left( \sqrt{\pi}v_{\mathrm{dr},e}/c_\mathrm{e\perp} \right)$ $ \mathrm{erfc}\!\left(-{v_{\mathrm{dr},e}}/{c_\mathrm{e\perp}}\right)$. We also include a secondary electron population, assumed to follow the Maxwellian distribution defined in Eq.~(\ref{fe_case1}). A background magnetic field of $B_0=9.6\times10^{-8} \mathrm{T}$ is utilized in the plasma model.


The instability in Case 3 was validated in Ref.~\onlinecite{xie2025efficient}, which demonstrated that the BO solver gives consistent real frequencies and growth rates for the analytical and fitted distributions.

Figure~\ref{fig:ring_beam_case} presents a comparison of the results obtained from BO and ALPS. The real frequencies and growth rates obtained from ALPS agree very well with the corresponding BO results throughout the entire wavenumber range.

\subsection{Case 4: Ion Cyclotron Emission Driven by an alpha Ring-Beam Distribution}

\begin{figure}[t!]
\centering
\includegraphics[width=0.95\columnwidth]{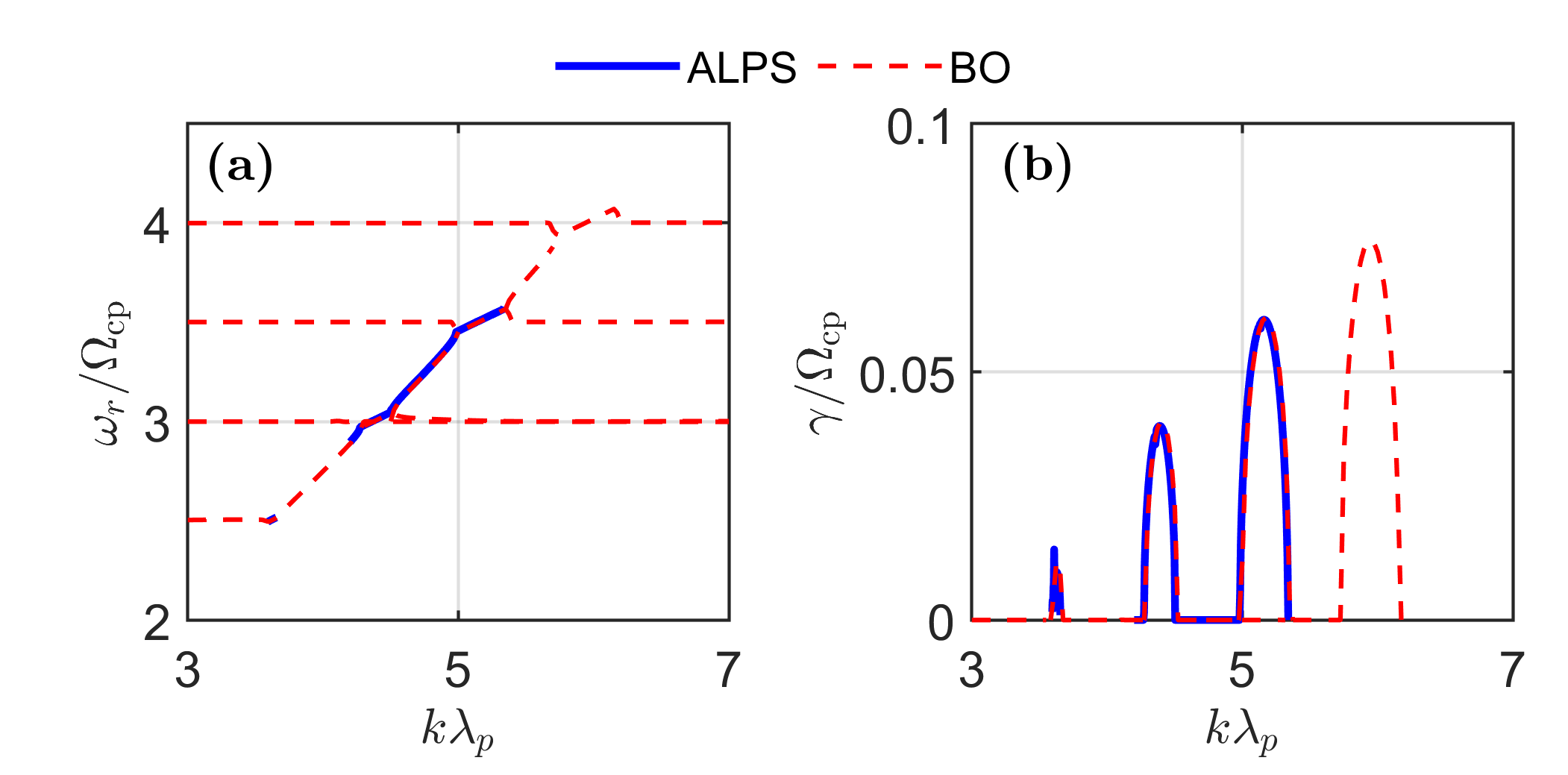}
\caption{Case 4: Comparison of ICE driven by an alpha ring-beam distribution. (a) Real frequency $\omega_r$ and (b) growth rate $\gamma$. 
The analytical expression for the alpha ring-beam distribution follows the same form as Eq.~(\ref{fe_case3}), with the particle specie replaced by alpha particle.
}
\label{fig:proton_ring_beam_case}
\end{figure}

In Case~4, we consider ion cyclotron emission (ICE) at $\theta=89.5^\circ$, driven by an alpha ring-beam distribution. Its analytical expression is identical to that in Eq.~(\ref{fe_case3}), but with the subscript ``$\alpha$'' instead of ``e''. Other two particle components--deuterium ions and electrons--are restricted to a Maxwellian distribution following Eq.~(\ref{fe_case1}). The background magnetic field is $B_0=2.1$~T.

This case has been verified in Ref.~\onlinecite{xie2025efficient}, in which the instability results from BO are consistent with those reported in Ref.~\onlinecite{warwick2018anomalous}.


The comparison between BO and ALPS for Case~4 is shown in Figure~\ref{fig:proton_ring_beam_case}. This figure shows only the first three growing modes with $\omega_r < 3.5\Omega_\mathrm{cp}$. The two solvers yield consistent real frequencies and growth rates for these modes. Additional tests confirm that ALPS can also calculate higher-order modes, with results that remain consistent with those from BO.

\subsection{Case 5: Quasi-perpendicular Instability Driven by a Proton Shell Distribution}

\begin{figure}[t!]
\centering
\includegraphics[width=0.95\columnwidth]{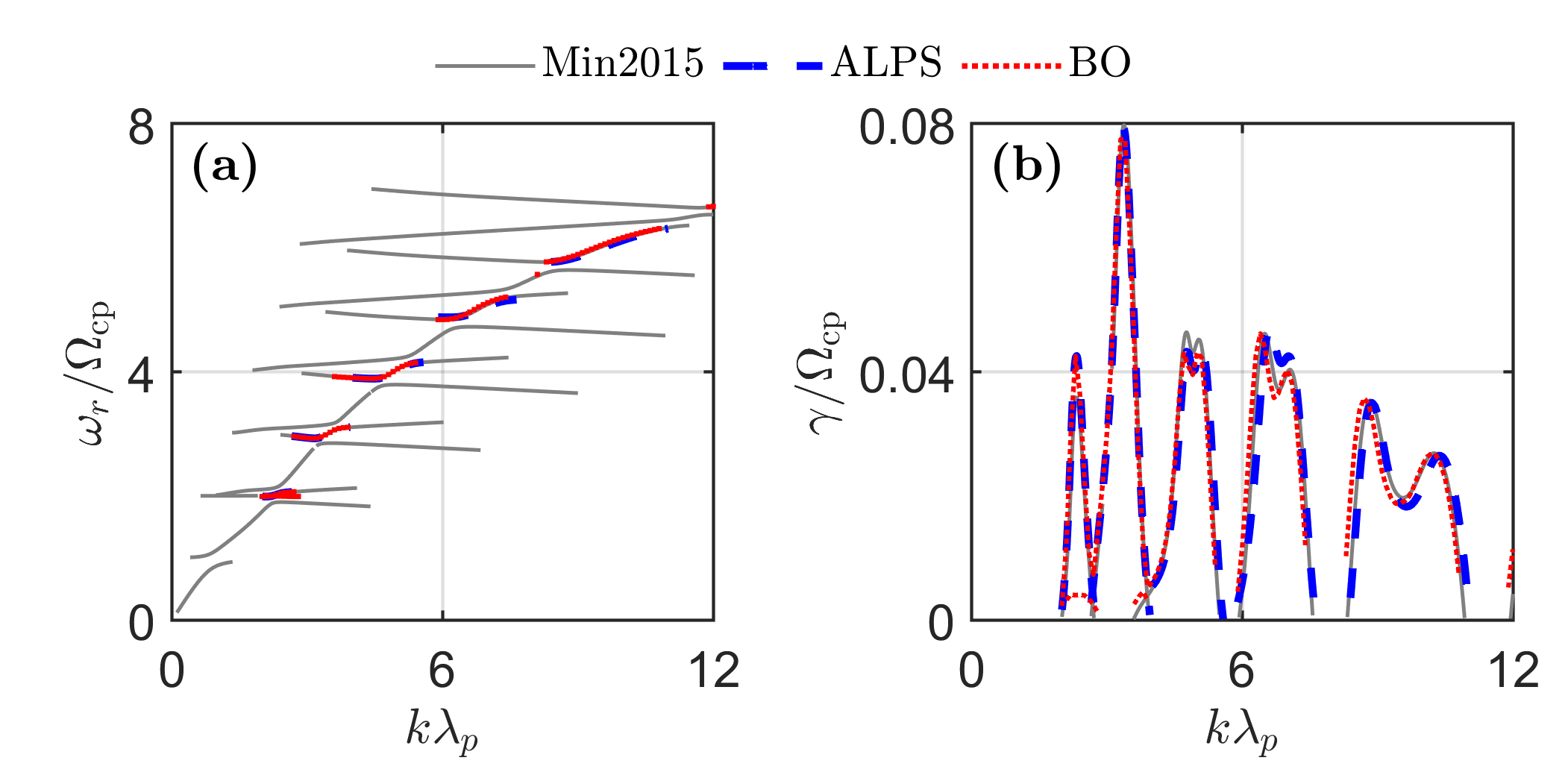}
\caption{Case 5: Comparison of the quasi-perpendicular instability driven by the proton shell distribution.  (a) Real frequency $\omega_r$ and (b) growth rate $\gamma$. The black, red, and blue curves represent the results from Ref.~\onlinecite{min2015fast}, BO, and ALPS, respectively. The analytical expression for the proton shell distribution is given in Eq.~(\ref{fp_case5}).}
\label{fig:shell_case}
\end{figure}

In Case~5, we consider a quasi-perpendicular instability driven by a proton shell distribution, given by the analytical expression \cite{min2015fast}:
\begin{equation}
\label{fp_case5}
f_{p}= \frac{1}{\pi^{3/2}c_\mathrm{p}^3A_{p}} \exp\!\left[-\frac{(v-v_{d,p})^2}{c_\mathrm{p}^2}\right],
\end{equation}
with $v=\sqrt{v_\parallel^2+v_\perp^2}$ and $A_{p}= ({2}/{\sqrt{\pi}}) ({v_{d,p}}/{c_\mathrm{p}})$ $ \mathrm{exp} \left( -v_{d,p}^2/c_\mathrm{p}^2 \right)+\left(2{v_{d,p}^2}/{c_\mathrm{p}^2}+1\right)\mathrm{erfc}\!\left(-{v_{d,p}}/{c_\mathrm{p}}\right)$. 

The electron population is assumed to be a Maxwellian distribution. The plasma parameters are summarized in Table~\ref{tab:particle_all_cases}. The background magnetic field is $B_0=3.28\times10^{-8}$~T, and the wave propagation angle is $\theta=89.5^\circ$.

The comparison of the quasi-perpendicular instability is shown in Figure~\ref{fig:shell_case}. The results from BO and ALPS are highly consistent with each other, and they both agree well with the results reported in Ref.~\onlinecite{min2015fast}.

\subsection{Case 6: Ion Cyclotron Waves Excited by a Proton Core-Beam Distribution}

\begin{figure}[h]
\centering
\includegraphics[width=0.95\columnwidth]{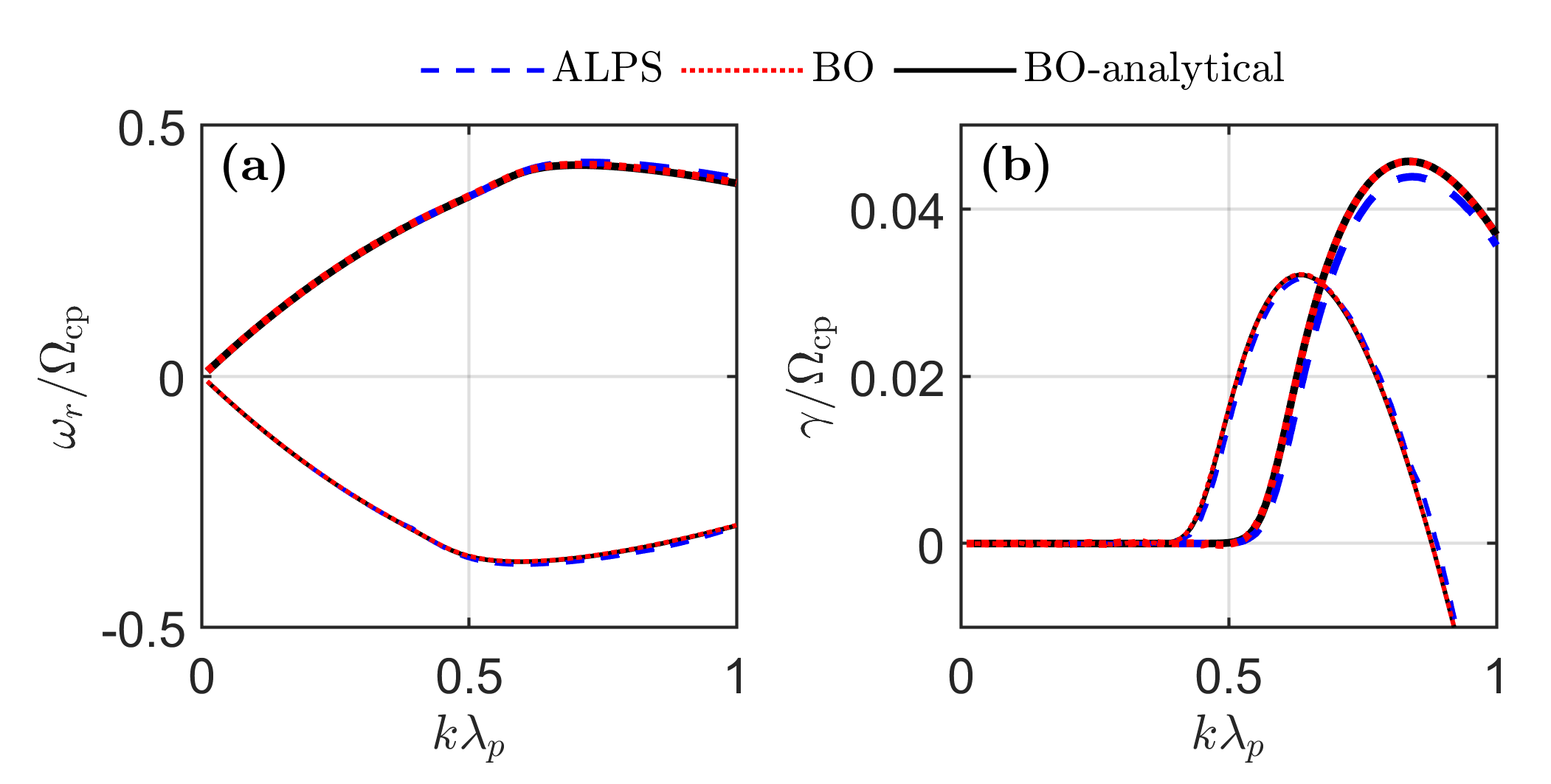}
\caption{Case 6: Comparison of counter-propagating ion cyclotron waves excited by a proton core-beam distribution. (a) Real frequency $\omega_r$ and (b) growth rate $\gamma$. The black, red, and blue curves represent the numerical results obtained using the BO analytical distribution, the BO fitted distribution, and the ALPS fitted distribution, respectively. 
}
\label{fig:core_beam_case}
\end{figure}

In Case 6, we consider a proton core-beam instability model motivated by Ref.~\onlinecite{shi2024ApJ}, while adopting plasma parameters chosen for the present comparison. The plasma model consists of three particle components---a proton core, a proton beam, and electrons---embedded in a background magnetic field of $B_0=7.50\times10^{-7}$~T. Both proton populations are assumed to follow Maxwellian distributions with perpendicular temperature anisotropy, whereas the electrons follow an isotropic Maxwellian distribution. The plasma parameters are summarized in Table~\ref{tab:particle_all_cases}.

The combined effects of proton temperature anisotropy and beam drift lead to the excitation of counter-propagating ion cyclotron waves with unequal growth rates, as shown in Fig.~\ref{fig:core_beam_case}. This figure also includes the numerical results obtained from the analytical distribution evaluated by BO. The results derived from the BO analytical and BO fitted distributions are highly consistent with each other, whereas those from the ALPS fitted distribution show slight deviations in the calculated growth rates.

\section{Instability Comparison for an Observational Proton VDF}\label{sec4}

\begin{figure}[t!]
\centering
\includegraphics[width=0.95\columnwidth]{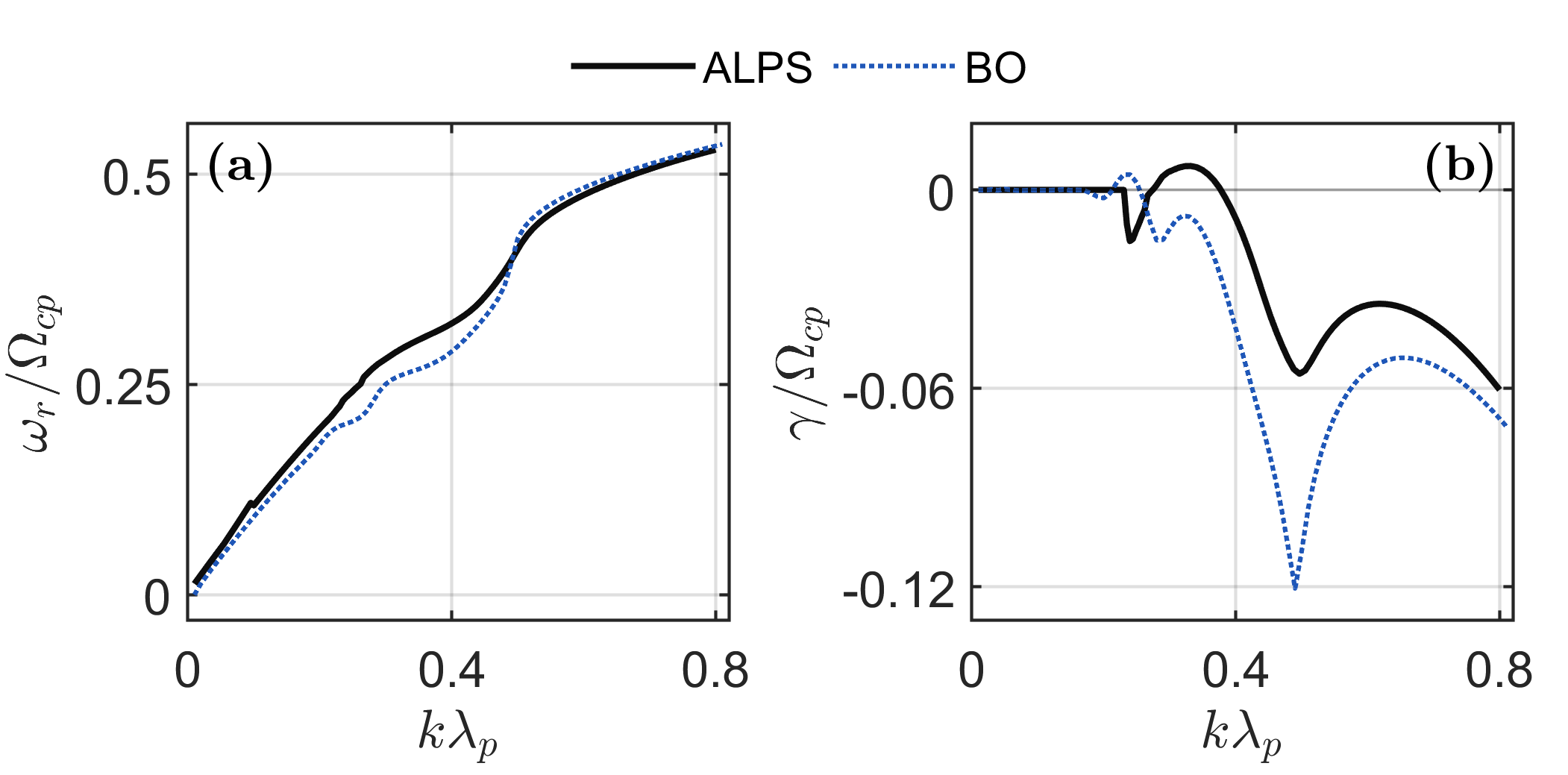}
\caption{Case 7: Comparison of the instability calculation results for an observational proton VDF. (a) The real frequencies and (b) the growth rates.}
\label{fig:observed_icw_case}
\end{figure}


To compare the two solvers using an observationally derived distribution, we consider a proton VDF measured in the terrestrial magnetosheath \cite{Zhao_2020}. This event is characterized by a pronounced perpendicular temperature anisotropy in the proton population.

Figures~\ref{fig:observed_icw_case}(a) and (b) compare the dispersion results obtained from BO and ALPS for the observed proton VDF. A polarization analysis indicates that the unstable mode is left-hand polarized, with an ellipticity of $-1$, consistent with the ICW observed during the event reported in Ref.~\onlinecite{Zhao_2020}.

For the real frequency shown in Figure~\ref{fig:observed_icw_case}(a), the ALPS result is broadly consistent with that from BO, although noticeable differences remain. For the corresponding growth rates shown in Figure~\ref{fig:observed_icw_case}(b), however, the two solvers show substantial quantitative differences. In particular, the unstable wavenumber ranges ($\gamma>0$) differ between the two calculations: the ALPS growth rate reaches its maximum at approximately $k \lambda_p=0.3$, whereas the BO result peaks near $k \lambda_p=0.23$. This discrepancy is mainly associated with differences in the fitting accuracy of the observed VDF, as discussed in Appendix~\ref{fit_com}, where the BO fit is shown to be less accurate than the ALPS fit.



It should be noted that a recent study \cite{Xie2026} showed that an initial-value approach is needed to solve the wave dispersion relation for noisy measured particle VDFs, because conventional dispersion-relation solvers are formally valid only in the asymptotic limit $t\to\infty$. This issue will be examined in detail in future work.

\section{Comparison of Solver Reliability and Computational Efficiency}

\begin{table*}[htbp]
\caption{Summary of Solver Accuracy and Runtime Across Benchmark Cases$^a$.}
\label{tab:solver_comparison}
\begin{ruledtabular}
\begin{tabular}{l c c c c}
Case & ALPS accuracy & ALPS runtime & BO accuracy & BO runtime \\
\hline
Case 1: Proton kappa distribution & Good & 5 min & Limited$^{b}$ & 1 min \\
Case 2: Product-kappa distribution & Good & 5 min & Limited$^{c}$ & 1 min \\
Case 3: Electron ring-beam distribution & Good & 30 min & Good & 5 min \\
Case 4: Alpha ring-beam distribution & Good & 30 min & Good & 5 min \\
Case 5: Proton shell distribution & Good & 30 min & Good & 5 min \\
Case 6: Proton core-beam distribution & Good & 5 min & Good & 1 min \\
Case 7: An observed distribution & Good & 10 min & Limited$^{d}$ & 1 min \\
\end{tabular}
\end{ruledtabular}
\vspace{1mm}
{\footnotesize $^{a}$ The listed runtimes were measured on an Intel Core Ultra 9 285K CPU; ALPS runtimes are for one target root, whereas BO runtimes are for all roots in one run. $^{b}$ The BO result is less smooth at $\kappa=4$. $^{c}$ BO underfits the distribution tail and underestimates the growth rate at $\kappa=3$. $^{d}$ BO fits poorly in the large velocity range.}
\end{table*}

\begin{table*}[htbp]
\caption{Qualitative Comparison Between BO and ALPS.}
\label{tab:qualitative_comparison}
{\renewcommand{\arraystretch}{1.2}
\begin{tabular}{l c c}
\hline
Feature & BO & ALPS \\
\hline
Treatment of externally imported VDFs & \shortstack[c]{Requires prior fitting\\or basis expansion.} & \shortstack[c]{Can operate directly on\\gridded distributions.} \\
\hline
Computational efficiency & Higher & Lower \\
\hline
Root-solution strategy & \shortstack[c]{Matrix-eigenvalue formulation;\\multiple roots per run;\\weak dependence on initial guesses.} & \shortstack[c]{Complex-plane iterative search;\\usually requires initial guesses\\or map guidance.} \\
\hline
Adaptability to complicated VDFs & Lower & Higher \\
\hline
Difficulty in determining fitting parameters & \shortstack[c]{Higher; manual tuning\\is often required.} & \shortstack[c]{Lower; usually no manual\\adjustment is needed.} \\
\hline
Relativistic treatment & Non-relativistic & Supports relativistic calculations \\
\hline
Recommended applications & \shortstack[c]{Rapid mode surveys when\\initial roots are unavailable.} & \shortstack[c]{High-accuracy analysis of\\complicated gridded distributions.} \\
\hline
\end{tabular}
}
\end{table*}

After comparing the instability results for the individual benchmark cases, we now summarize the reliability and computational cost of the two solvers.

A case-by-case comparison of BO and ALPS is presented in Table~\ref{tab:solver_comparison}. For each case, the wave frequencies were evaluated using approximately 50--300 sampled wavenumber points, depending on the wavenumber range over which the unstable modes were traced. Fewer than 100 points were sufficient for cases with a narrow wavenumber range, such as the proton kappa case, whereas up to about 300 points were used for broader cases, such as the proton shell case.

The two solvers show different strengths in accuracy and computational cost. ALPS gives reliable results for all benchmark cases considered here, whereas BO shows reduced accuracy for the two low-kappa cases and for the observed proton distribution. These limitations mainly arise from inaccurate fitting of the corresponding particle VDFs in the BO implementation. In contrast, BO is computationally more efficient: its runtimes are typically on the order of 1--5 min, compared with 5--30 min for ALPS. The higher computational cost of ALPS is due to its two-step procedure, which includes a preliminary scan in the $\omega_r$-$\gamma$ plane to identify suitable initial guesses, followed by a scan over $k$. This additional search becomes particularly expensive when multiple branches need to be traced.

The practical differences between BO and ALPS are further summarized in Table~\ref{tab:qualitative_comparison}. BO requires prior fitting or basis expansion for externally imported VDFs, whereas ALPS can operate directly on gridded distributions. This difference affects their applicability to complicated VDFs: BO is more sensitive to the choice of fitting parameters and often requires manual tuning, while ALPS generally provides a more direct treatment of complex or observed distributions. In terms of the root-solution strategy, BO uses a matrix-eigenvalue formulation and can obtain multiple roots in a single run, with relatively weak dependence on initial guesses. This makes BO computationally efficient and useful for rapid mode surveys, especially when the relevant roots are not known a priori. By contrast, ALPS solves the dispersion relation through an iterative search in the complex-frequency plane, which usually requires initial guesses or guidance from an $\omega_r$-$\gamma$ map. As a result, ALPS is generally slower, but it is better suited for high-accuracy analysis of complicated gridded VDFs and also supports relativistic calculations. These comparisons indicate that BO and ALPS have complementary strengths: BO is advantageous for efficient searches over multiple wave branches, whereas ALPS is more appropriate when accurate treatment of complex or observed VDFs is required.

In addition, recent work find that the computational cost of BO can be further reduced by exploiting its matrix structure, leading to a speedup of roughly two orders of magnitude; this improvement will be examined in future work \cite{Tian2026}.

\section{Summary and Discussion}\label{sec5}


This study compares the instability results obtained from the BO and ALPS solvers for arbitrary particle velocity VDFs. The two solvers show good performance and mutual agreement across the six benchmark cases with analytical expressions, although clear limitations of BO are found in specific cases. In particular, BO has difficulty accurately resolving kappa and product-kappa distributions with low $\kappa$ indices, while it gives reliable results for the other tested analytical VDFs. These limitations mainly arise from the BO-HH fitting scheme, which becomes less accurate for distributions extending over a broad velocity range. This fitting error appears to be the primary cause of the unreliable instability results obtained for the low-$\kappa$ cases. By contrast, ALPS provides reliable results for all analytical VDFs tested in this study.


We also compared the excitation of ion cyclotron waves using an observed proton VDF. Although the two solvers yield roughly consistent real frequencies for the counter-propagating ion cyclotron waves, their calculated growth rates differ significantly. This discrepancy mainly arises from the inadequate BO-HH fitting of the observed VDF. The fitting difficulty is related to the orthogonal-basis expansion used in BO-HH. For distributions spanning a wide dynamic range or containing fine-scale structures, the basis expansion may become numerically ill-conditioned under double-precision arithmetic. As a result, some features of the observed VDF may not be accurately represented, leading to errors in the calculated growth rates. In Appendix~\ref{app:bo_parameter_sensitivity}, we provide further details on how reliable BO fitting can be obtained.

From the viewpoint of accuracy, ALPS gives more reliable results in the present benchmark tests. For growing roots with $\gamma>0$, ALPS calculates the instability directly on the distribution grid. For damped roots with $\gamma<0$, ALPS can also provide a more accurate fitted representation of the distribution. Therefore, ALPS can consistently provide reliable results for both growing and damped roots. Its main limitation is computational cost. ALPS usually requires a preliminary scan of the dispersion map of $\log |\det \mathbf{D}(\omega_r,\gamma)|$ to identify suitable initial guesses for the target roots, and this step can be time-consuming. The convergence of the iterative solver also affects the total runtime. If the initial guesses are not sufficiently close to the target roots, or if the tracked branch passes through regions where root jumping occurs, the iteration may converge more slowly or require additional adjustments.

Because BO uses a matrix-eigenvalue formulation, its main advantage is that it can obtain all wave modes at a given wave vector in a single run. This feature makes BO useful for rapid surveys of unstable modes, especially when the relevant branches are not known in advance. ALPS, on the other hand, relies on iterative root finding in the complex-frequency plane and is better suited for refined calculations once the target modes have been identified. A practical strategy is therefore to use BO for an initial broad search of unstable branches and then use ALPS for more accurate evaluation of selected modes. This combined approach can improve both mode identification and the reliability of the final dispersion results.



The MATLAB version source code for BO is available at \url{https://github.com/hsxie/boarbitrary}. The Julia version of BO, PlasmaBO.jl, is available at \url{https://github.com/JuliaSpacePhysics/PlasmaBO.jl}. The ALPS code is available at \url{https://github.com/danielver02/ALPS}.

\begin{acknowledgments}
This study is supported by NSFC 42374196 and Scientific Research Foundation of Hunan provincial Education Department grant 24B0320. The ALPS project received support from UCL's Advanced Research Computing Centre through the Open Source Software Sustainability Funding scheme. D.V. is supported by STFC Consolidated Grant ST/W/001004/1. This research was supported by the International Space Science Institute (ISSI) in Bern, through ISSI International Team project \#612 (Excitation and Dissipation of Kinetic-Scale Fluctuations in Space Plasmas) led by K.~G.~Klein.
\end{acknowledgments}

\appendix
\section{Plasma Parameters for the Six Cases}
\label{app_parameters}

The plasma parameters used in the six cases are summarized in Table~\ref{tab:particle_all_cases}.

\begin{table*}[htbp]
\caption{Plasma parameters for all benchmark distribution cases.}
\label{tab:particle_all_cases}
\begin{ruledtabular}
{\renewcommand{\arraystretch}{1.15}
\begin{tabular}{l c c c c c c c}
Case & $q_{s}(e)$ & $m_{s}(m_{\mathrm{unit}})$ & $n_{s}(\mathrm{m}^{-3})$ & $T_{\mathrm{zs}}(\mathrm{eV})$ & $T_{\mathrm{ps}}(\mathrm{eV})$ & $v_{\mathrm{dsz}}/c$ & $v_{\mathrm{dsr}}/c$ \\
\hline
\multirow{2}{*}{Case 1: Proton kappa distribution}
& $1$ & $1.0$ & $5.0\times10^{19}$ & $1986.734$ & $993.367$ & $0.0$ & $0.0$ \\
& $-1$ & $5.447\times10^{-4}$ & $5.0\times10^{19}$ & $496.683$ & $496.683$ & $0.0$ & $0.0$ \\
\hline
\multirow{2}{*}{Case 2: Product-kappa distribution}
& $1$ & $1.0$ & $1.0\times10^{7}$ & $50$ & $50$ & $0.0$ & $0.0$ \\
& $-1$ & $5.447\times10^{-4}$ & $1.0\times10^{7}$ & $102$ & $408$ & $0.0$ & $0.0$ \\
\hline
\multirow{2}{*}{Case 3: Electron ring-beam distribution}
& $-1$ & $5.447\times10^{-4}$ & $1.0\times10^{5}$ & $51$ & $51$ & $0.1$ & $0.05$ \\
& $-1$ & $5.447\times10^{-4}$ & $9.0\times10^{5}$ & $51$ & $51$ & $0$ & $0$ \\
\hline
\multirow{3}{*}{Case 4: Alpha ring-beam distribution}
& $2$ & $4$ & $1.0\times10^{16}$ & $1.0\times10^{3}$ & $1.0\times10^{3}$ & $0$ & $0.045$ \\
& $1$ & $2$ & $9.98\times10^{18}$ & $1.0\times10^{3}$ & $1.0\times10^{3}$ & $0$ & $0$ \\
& $-1$ & $5.447\times10^{-4}$ & $1.0\times10^{19}$ & $1.0\times10^{3}$ & $1.0\times10^{3}$ & $0$ & $0$ \\
\hline
\multirow{3}{*}{Case 5: Proton shell distribution}
& $1$ & $1$ & $0.5\times10^{5}$ & $1\times10^{3}$ & $1\times10^{3}$ & $0$ & $0.133^{*}$ \\
& $1$ & $1$ & $4.5\times10^{5}$ & $10$ & $10$ & $0$ & $0$ \\
& $-1$ & $1\times10^{-2}$ & $5\times10^{5}$ & $10$ & $10$ & $0$ & $0$ \\
\hline
\multirow{3}{*}{Case 6: Proton core-beam distribution}
& $1$ & $1$ & $2.53\times10^{9}$ & $20$ & $100$ & $-2.87\times10^{-4}$ & $0$ \\
& $1$ & $1$ & $2.17\times10^{9}$ & $48$ & $170$ & $3.33\times10^{-4}$ & $0$ \\
& $-1$ & $5.447\times10^{-4}$ & $4.7\times10^{9}$ & $50$ & $50$ & $0$ & $0$  \\
\end{tabular}
}
\end{ruledtabular}
\vspace{1mm}
{\footnotesize $^{*}$ For the shell distribution, $v_{\mathrm{dsr}}$ denotes $v_{d,p}$ in the analytical expression and thus differs physically from the perpendicular drift speed used in the other cases.}
\end{table*}

\section{Comparison of Fitting Results}
\label{fit_com}

This appendix compares the VDF representations used by ALPS and BO. As shown in Secs.~\ref{sec3} and \ref{sec4}, the two solvers give consistent instability results for most benchmark cases, but definitive differences appear for several distributions. These differences are closely related to the fitting accuracy of input VDFs.

\begin{figure}[t]
\centering
\includegraphics[width=1\columnwidth]{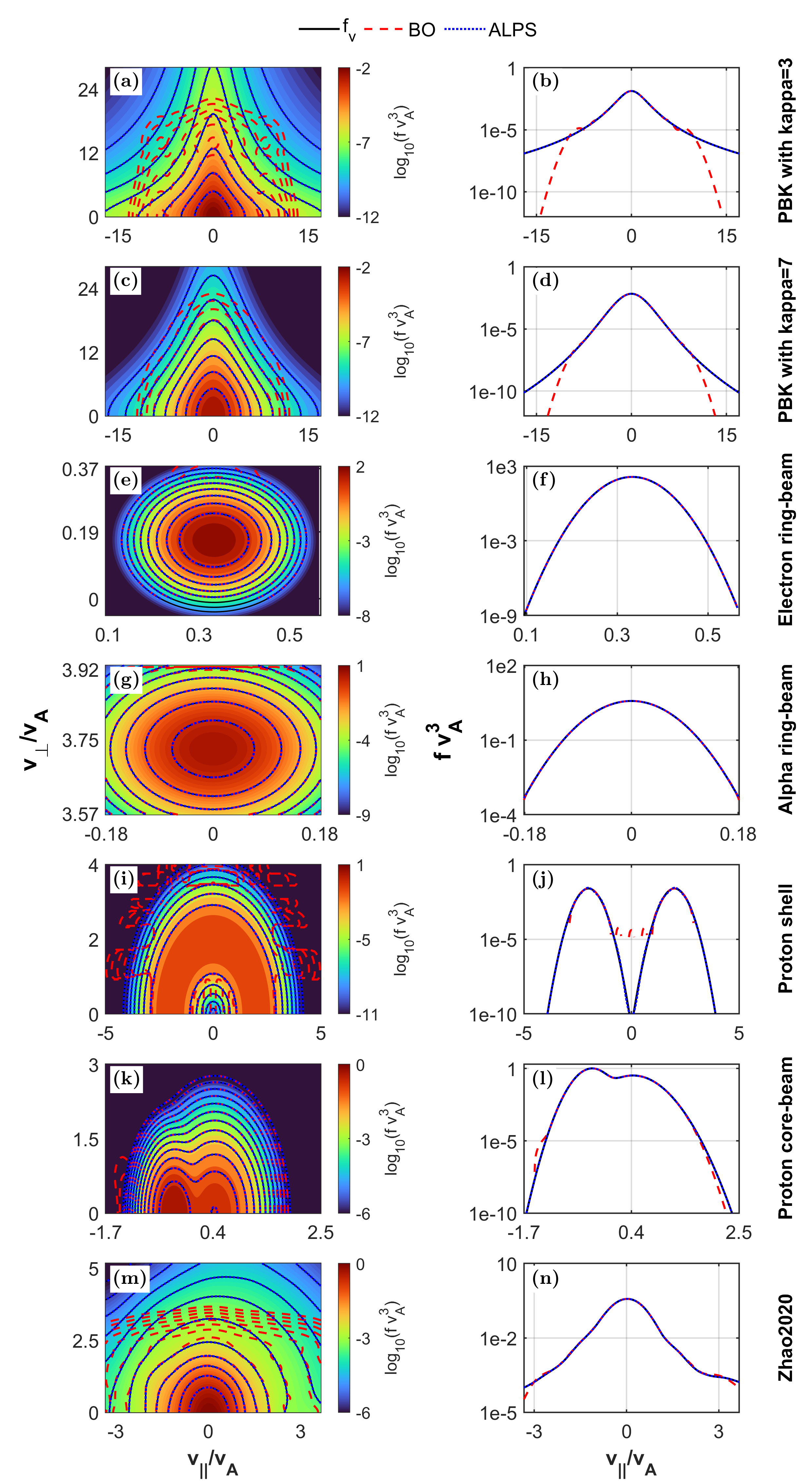}
\caption{Comparison of the VDF fits obtained from BO and ALPS. The left column shows the fits in two-dimensional velocity space, where the black solid contours denote the original distribution, the red dashed contours denote the BO-fitted result, and the blue dotted contours denote the ALPS-fitted result. The right column shows one-dimensional cuts at the value of $v_\perp$ corresponding to the maximum of each distribution; the line styles are the same as those used for the contours in the left column. Panels (a, b) show the PBK distribution with $\kappa=3$; (c, d) the PBK distribution with $\kappa=7$; (e, f) the electron ring-beam distribution; (g, h) the alpha ring-beam distribution; (i, j) the proton shell distribution; (k, l) the proton core-beam distribution; and (m, n) the observationally derived VDF from Ref.~\onlinecite{Zhao_2020}.}
\label{fig:fit_compare_all}
\end{figure}

Figure~\ref{fig:fit_compare_all} compares the original and fitted distributions for all analytical and observationally derived cases. We use the Alfv\'en speed of the reference species $v_A$ for velocity normalization. Specifically, the electron Alfv\'en speed is used for the product-bi-kappa and electron ring-beam distributions, whereas the proton Alfv\'en speed is used for the alpha ring-beam, shell, core-beam, and observational distributions.

Figure~\ref{fig:fit_compare_all} shows that ALPS preserves the original distribution more accurately than BO. BO exhibits more conisistent fitting for the high-$\kappa$ PBK, electron ring-beam, and alpha ring-beam distributions, but show larger discrepancies for the low-$\kappa$ PBK, shell, and observational distribution.

\section{Sensitivity of BO-HH Results to Expansion Parameters}
\label{app:bo_parameter_sensitivity}

\begin{figure}[htbp]
\centering
\includegraphics[width=0.45\textwidth]{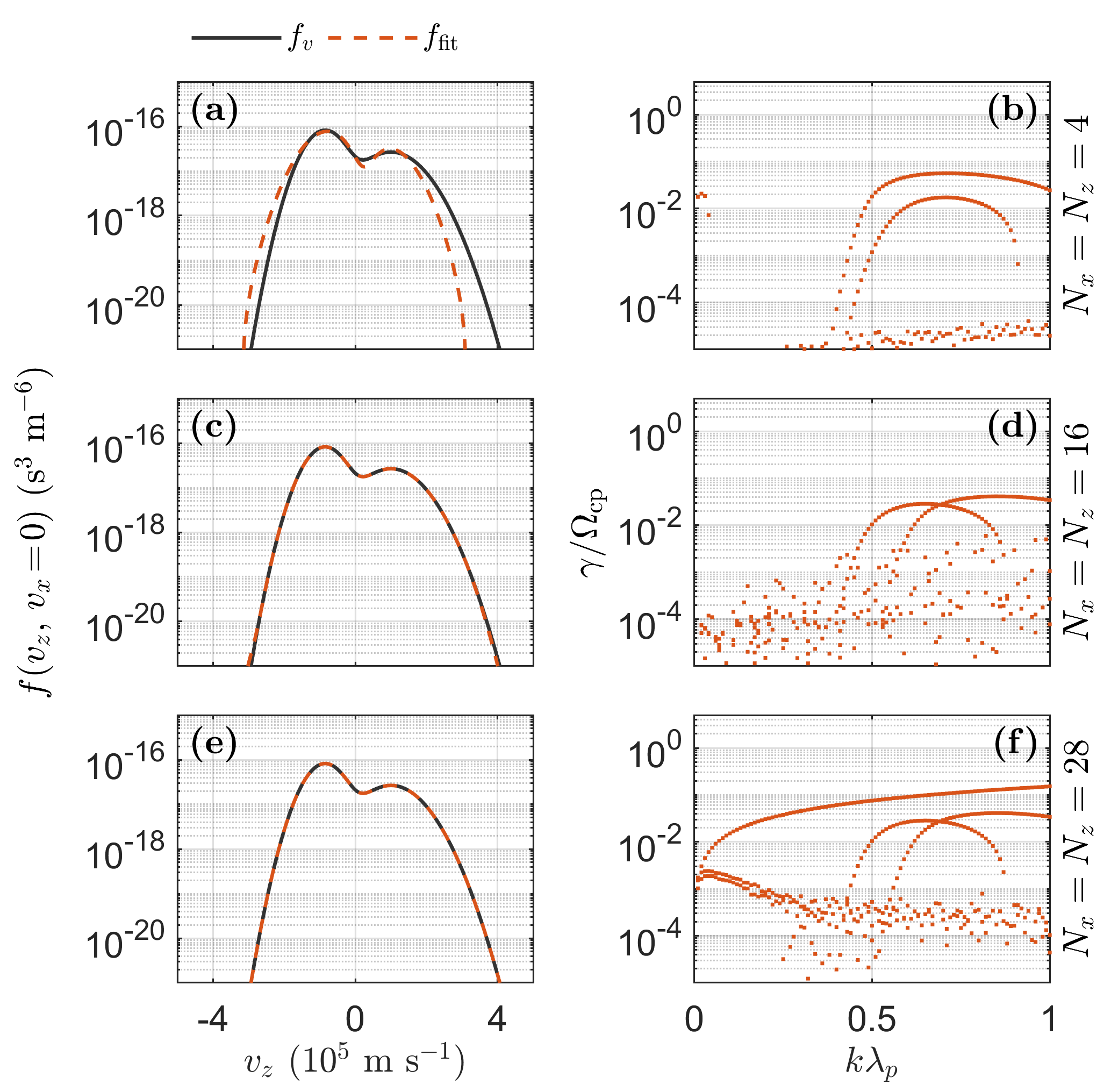}
\caption{Effect of the Hermite expansion order on the BO-HH fitting and the resulting roots. Panels (a)--(c) compare the BO-HH fitted distributions with the original distribution for $N_x=N_z=4$, $16$, and $28$, respectively. Panels (d)--(f) show the roots calculated by BO-HH for the corresponding choices of the expansion parameters.}
\label{fig:bo_parameter_sensitivity}
\end{figure}

During the BO-HH fitting process, the fitted VDF can exhibit inaccurate velocity-space derivatives ($\partial f/\partial v$), especially in the high-velocity tail regions. These spurious gradients may generate nonphysical roots. Consequently, the Hermite-polynomial expansion orders ($N_x$ and $N_z$) and the dispersion tensor expansion order ($J$) must be chosen carefully to ensure a reliable instability analysis.

Figure~\ref{fig:bo_parameter_sensitivity} illustrates this sensitivity by varying $N_x$ and $N_z$ at fixed $J$. We fit the core-beam distribution using three different values of $N_x$ and $N_z$, while keeping the dispersion tensor expansion order fixed at $J=24$. As shown in Figures~\ref{fig:bo_parameter_sensitivity}(a)--(c), the fitting accuracy improves as $N$ increases, with $N_x=N_z=28$ providing the closest visual agreement with the original distribution.

However, Figures~\ref{fig:bo_parameter_sensitivity}(d)--(f) show that a visually better VDF fit does not necessarily lead to more reliable instability roots. The case with $N_x=N_z=16$ recovers the two reference instabilities most accurately. In contrast, when the expansion order is too low ($N_x=N_z=4$), the fitted distribution is insufficiently accurate and the resulting roots fail to reproduce the reference instabilities. When the expansion order is too high relative to $J$ ($N_x=N_z=28$ with $J=24$), the two physical instabilities are still recovered, but two additional spurious instabilities also appear.

This test indicates that an intermediate range of expansion orders is required. The Hermite expansion order must be large enough to reproduce the main structure of the VDF, but not so large that the fitted distribution introduces artificial fine-scale features or derivative errors. As a practical guideline, we recommend choosing $N_x$ and $N_z$ sufficiently large to fit the main VDF structure while keeping them lower than the dispersion tensor expansion order, for example with $J-N\ge4$ when $N_x=N_z=N$.

\bibliographystyle{aipnum4-1}
\clearpage
\bibliography{aipsamp}

\end{document}